\documentclass[10pt,aps,prb,superscriptaddress,twocolumn,a4paper,floatfix]{revtex4-2}
\usepackage{amsmath}%
\usepackage{svg}
\usepackage{amsfonts}%
\usepackage{amssymb}%
\usepackage[latin1]{inputenc}
\usepackage{graphicx}
\usepackage{color}
\usepackage[english]{babel}
\usepackage{natbib}
\usepackage[colorlinks=true, citecolor=blue, linkcolor=blue, urlcolor=blue]{hyperref}
\usepackage{booktabs}
\usepackage{enumitem}
\usepackage{subfigure}
\usepackage{multirow}
\usepackage{dsfont}
\usepackage{braket}
\usepackage{dcolumn}
\usepackage{bm}% bold math
\usepackage[parse-numbers=true]{siunitx}
\usepackage{textgreek}
\usepackage{wasysym}
\usepackage{tabularx}
\usepackage{float}
\usepackage{subfloat}
\usepackage{ragged2e}
\fontencoding{T1}\selectfont

\usepackage[capitalize]{cleveref}

\usepackage{comment}

\graphicspath{{images_colors/}}

\begin{document}

\title{Experimental Realization of an Acoustic Kagome-Model Simulator}

\title{Acoustic Realization of the kagome Lattice}

\title{Acoustic Probing of Sublattice Physics in a kagome Lattice}

\title{Acoustic Wave-Function Imaging of Sublattice Physics in a Kagome Lattice}

% \title{Disentangling Band Structure and Topology in an Acoustic kagome Lattice}

\author{L.~M\"{u}ller}
\thanks{These authors contributed equally to this work.}

\author{N.~Endres}
\thanks{These authors contributed equally to this work.}

\author{B.~Geldiyev}

\affiliation{Experimentelle Physik VII and W\"{u}rzburg-Dresden Cluster of Excellence ctd.qmat, Julius-Maximilians-Universit\"{a}t W\"{u}rzburg, Am Hubland, D-97074 W\"{u}rzburg, Germany}

\author{S.~Widmann}

\affiliation{Technische Physik and W\"{u}rzburg-Dresden Cluster of Excellence ctd.qmat, Julius-Maximilians-Universit\"{a}t W\"{u}rzburg, Am Hubland, D-97074 W\"{u}rzburg, Germany}

\author{K.~Burkard}

\author{M.~\"{U}nzelmann}
	
\author{F.~Reinert}
\email[corresponding author: ]{reinert@physik.uni-wuerzburg.de}

\affiliation{Experimentelle Physik VII and W\"{u}rzburg-Dresden Cluster of Excellence ctd.qmat, Julius-Maximilians-Universit\"{a}t W\"{u}rzburg, Am Hubland, D-97074 W\"{u}rzburg, Germany}

\date{\today}

\renewcommand{\figurename}{FIG.}

\begin{abstract}

The canonical kagome band structure hosts a Dirac cone, saddle-point van Hove singularities (vHS), and a flat band. Characteristically, these features exhibit distinct sublattice localization within the three-site basis of the kagome lattice. However, experimental proof of this sublattice character is challenging and requires direct access to both the real-space wave function and its associated momentum-space texture. Here, we implement an acoustic kagome lattice on the cm scale and measure the full excitation spectrum, thereby imaging the wave function at each lattice site, including both amplitude and phase. This enables investigation of the momentum-dependent sublattice texture across the complete band structure, particularly at the flat band and at the mixed- and pure-sublattice van Hove singularities. Our results are in excellent agreement with minimal tight-binding model calculations and constitutes the first direct experimental observation of the sublattice-resolved band structure in a kagome lattice.
 
\end{abstract}


\maketitle

Research into kagome systems has intensified recently, driven by the discovery of topological kagome metals such as the $A \mathrm{V}_3 \mathrm{Sb}_5 ~ (A = \mathrm{K,\,Rb,\,Cs})$ family \cite{Ortiz2019} and others \cite{Ye2018, Liu2018, Kang2019, DiSante2026}. These materials exhibit diverse electronic and magnetic phenomena --- ranging from frustrated magnetism \cite{Kida2011, Meschke2021} and the anomalous Hall effect \cite{Lachman2020, Chen2021, Song2024} to unconventional superconductivity \cite{Ortiz2020, Wu2021, Guguchia2023, Holbaek2023} and density wave orders \cite{Luo2022, Teng2022, Cao2023, Park2025, Yu2012} --- and have thus come into the spotlight of both experimental and theoretical investigations \cite{Denner2021, Jiang2021, Mielke2022}. 
\\
The electronic structure in kagome quantum materials manifests itself in the characteristic three-site sublattice structure. The flat band arises, for instance, from destructive interference between two sublattice-hopping paths, which yields completely localized electrons. Moreover, the momentum-space sublattice wave function textures at the saddle-point vHS are supposed to alter emergent electron-correlation effects decisively \cite{Kiesel2012, Kiesel2013, Wang2013, Schwemmer2024}.
\\
Although all of this is theoretically well established, it is challenging to provide direct experimental evidence of the comprehensive sublattice fingerprint, for several reasons: 
First of all, the structural and multi-orbital complexity, inherent to almost all electronic kagome systems, often obscures the underlying physics. In many material candidates, the atomic two-dimensional (2D) kagome lattice is embedded within a complex 3D crystalline environment, making it difficult to isolate a purely planar, single-orbital physical picture \cite{Yin2022, Teng2023}.
Second, the full sub-unit-cell wave function information is generally difficult to address in electronic systems; the electronic wave function is \textit{not} a quantum mechanical observable.  For instance, angle-resolved photoelectron spectroscopy (ARPES), mostly used for band structure measurements of electronic systems, is inherently limited by its insensitivity to the phase of the Bloch electrons, i.e., it fails to capture the destructive interference and sublattice character defining the FB and vHS \cite{Kiesel2012, Kiesel2013, Wang2013, Schwemmer2024}.

%Metamaterials offer an experimental designer platform where the Hamiltonian is explicitly encoded into the lattice geometry \cite{Torrent2012, Yang2015, SerraGarcia2018, Peterson2018, Li2018, Wen2019, Milicevic2019, Huang2020, Wei2021, Karki2023, Zhang2019}. 
%\textcolor{red}{Metamaterials offer an experimental designer platform where the Hamiltonian is explicitly encoded into the lattice geometry and can be realized in many ways, e.g. photonic \cite{Zheludev2016, Milicevic2019}, phononic \cite{Mousavi2015, Li2018, Huang2020}, mechanical \cite{Florijn2014, Surjadi2019} and acoustic  \cite{Zhang2019, Wen2019, Wei2021} systems and topoelelectrical circuits \cite{Ningyuan2015, Imhof2018}.}
Metamaterials offer an experimental designer platform where the Hamiltonian is explicitly encoded into the lattice geometry \cite{Wang2009, Zheludev2016, Milicevic2019, Lu2014, Mousavi2015, Li2018, Huang2020, Florijn2014, Suesstrunk2015, Huber2016, Surjadi2019, Zhang2019, Wen2019, Wei2021, Ningyuan2015, Imhof2018, Lee2018, Klembt2018, Harder2021}.
The structural transparency provides a level of control rarely found in electronic systems, circumventing the need for complex materials chemistry to engineer an ideal kagome band structure. Furthermore, site-resolved measurements reveal the wave function localization and sublattice characteristics that remain elusive in electronic systems. In essence, this platform creates a one-to-one link between the physical structure and theoretical models, allowing for the unambiguous identification of the destructive interference and sublattice contribution that drive these characteristic kagome lattice features \cite{Ni2023, Zhu2023}.

\iffalse
This entanglement acts as a significant hindrance to resolving the defining features of the kagome band structure: the flat band (FB), resulting from destructive interference; the Dirac cone, characterized by linear dispersion; and the van Hove singularities (vHSs) that manifest specific sublattice characters. These hallmarks drive the electronic correlations and topological characteristics observed in real materials. 

However, their experimental identification remains a challenge, as sub-unit-cell information is generally difficult to address in electronic systems. For instance, angle-resolved photoelectron spectroscopy (ARPES) mostly used for band structure measurements of electronic systems, it is inherently limited by its insensitivity to the phase of the Bloch electrons, i.e., it fails to capture the destructive interference and sublattice character defining the FB and vHSs \cite{Kiesel2012, Kiesel2013, Wang2013, Schwemmer2024}.

% sublattice character through matrix element effects - Phase of the wavefunction---which is not a measurable observable in Photoemission

\fi

% the destructive interference and sublattice localization defining the kagome FB and vHS - such as multi-orbital hybridization or 3D interlayer coupling

In this Letter, we investigate the classical wave analog of an electronic kagome lattice using an acoustic metamaterial. While earlier works on metamaterials report observations including topological edge states \cite{Ni2017}, higher-order topological insulators \cite{Xue2019, Xiang2019}, and the non-Hermitian skin effect \cite{Zhong2025}, these studies do not address the sublattice-resolved band characteristics. In addition to recovering the kagome band structure \cite{Jiang2021_0, Xia2025}, we provide the hitherto elusive experimental access to the sublattice texture long predicted to govern its electronic structure \cite{Kiesel2012, Kiesel2013, Schwemmer2024}. 
\\
\\
\begin{figure*}
    \includegraphics[width=1\textwidth]{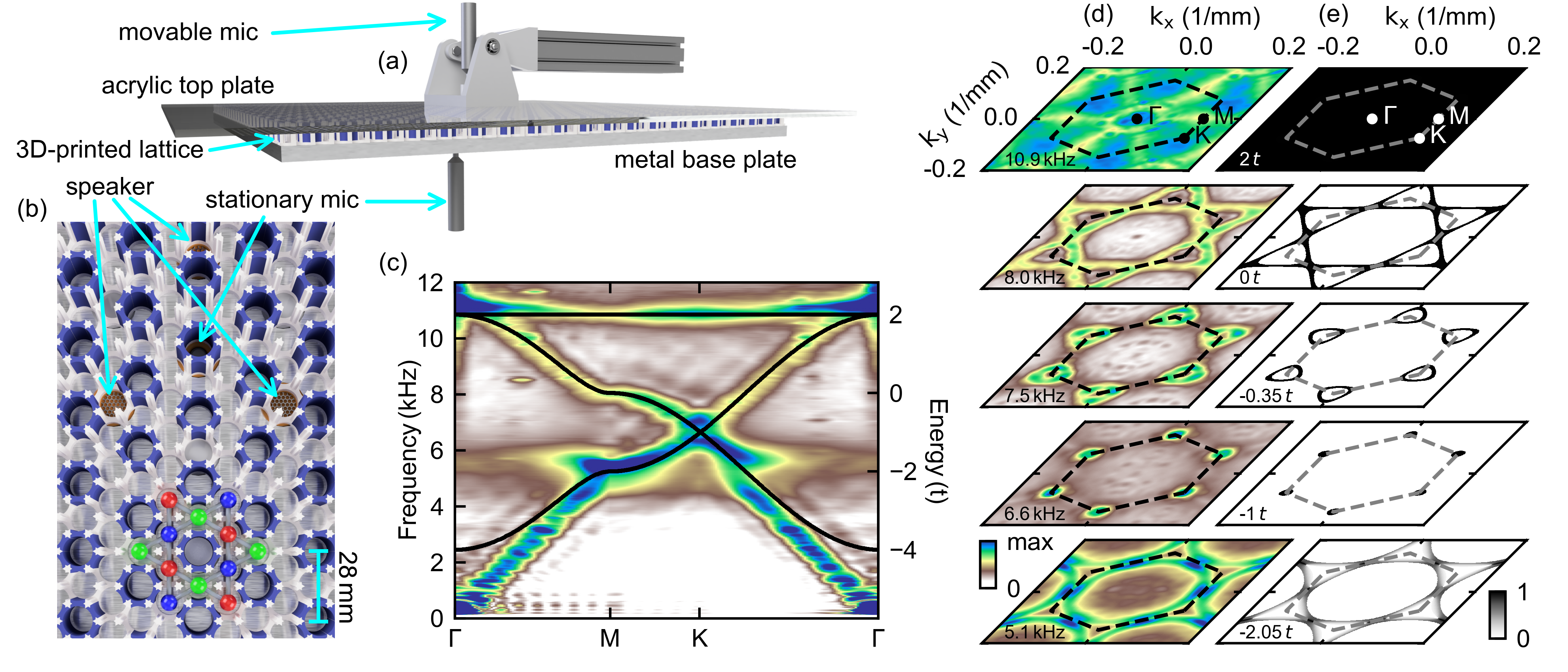}
    \justifying
    \caption{Acoustic kagome lattice and band structure. (a) Three-dimensional model of the experimental setup. (b) Top-view image of the 3D-printed kagome lattice. (c) Measured acoustic band structure, obtained by an incoherent sum of the response over all sublattices, overlaid with the minimal tight-binding (TB) model calculation (solid lines). (d) Experimental isofrequency surfaces at the selected frequencies (see inset). (e) Calculated isofrequency surfaces from the TB model including a non-zero hopping to the hollow site.}
    \label{fig:figure1}
\end{figure*}
The kagome lattice is realized here by a three-resonator basis (sublattices A, B, C) surrounding a central hollow site (H) (see \cref{fig:figure1}(b)). The structure is fabricated via 3D printing and consists of triangular pillars. The hollow sites are separated by 3D-printed walls (blue in \cref{fig:figure1}(b)) that are inserted between the pillars, resulting in a lattice constant of $a=28\,$mm. The lattice is bonded to an aluminum base plate featuring threaded holes for a speaker and a microphone. A second microphone is attached to a robotic arm for automated lateral scanning (see \cref{fig:figure1}(a,b)). The speaker has an aperture size of $1\,\mathrm{mm}$ and is fixed at a site of a selected sublattice. The robotic arm sequentially scans the lattice, and for each resonator the speaker emits a sinusoidal excitation signal whose frequency is swept linearly from $100$ to $12000\,\mathrm{Hz}$ over $30\,\mathrm{s}$. This procedure yields the site-resolved complex acoustic response, providing the amplitude and phase as a function of frequency. The measurements are repeated with the speaker positioned on the two remaining sublattices. This maximizes the spatial overlap between the driving field and the lattice eigenmodes, enabling access to a broader set of modes and accurately capturing the system's symmetry.

% \textcolor{red}{(this facilitates largest possible eigenvectors) ---} \textcolor{red}{The experiment is performed multiple times with the speaker at distinct lattices sites, facilitating an excitation of the largest possible amount of eigenvectors by maximizing spatial overlap.}

% To map the acoustic response

% via a $1$-$\mathrm{mm}$-diameter aluminum aperture, flush-mounted within the base plate

% \textcolor{red}{---a corner-sharing triangular network---} / captured \textcolor{red}{implemented/realized}

\begin{figure}
    \centering
    \includegraphics[width=\linewidth]{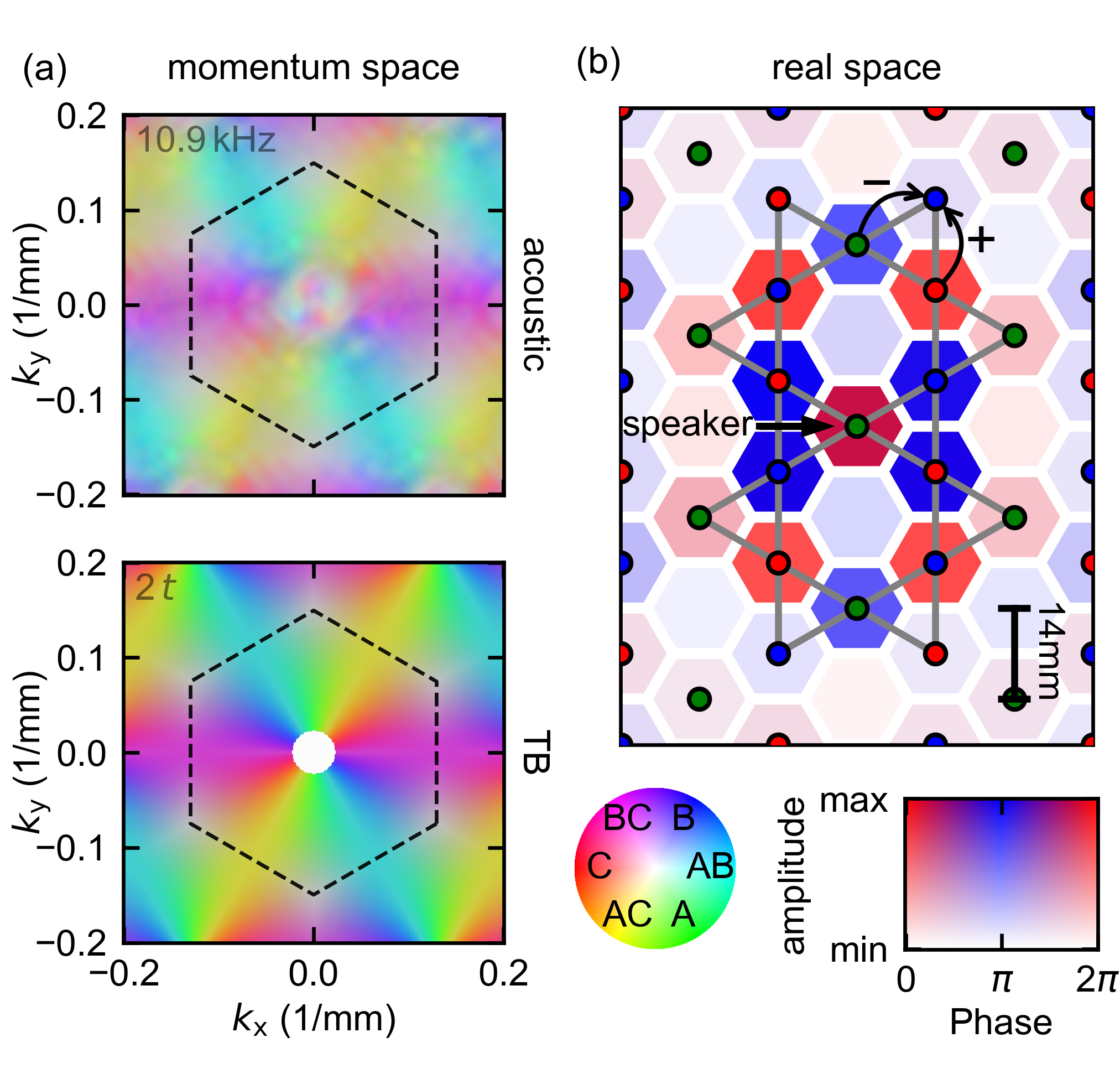}
    \caption{(a) A comparison of the flat band sublattice characteristics from experiment (top) and the TB model (bottom). The isofrequency contours reveal complex direction-dependent m- and p-type sublattice contributions. (b) Real-space phase and amplitude map at the flat-band frequency. The map demonstrates sublattice interference on the kagome net, characterized by a $\pi$ phase difference between adjacent resonators and vanishing amplitude at the hollow sites.}
    \label{fig:flat-band}
\end{figure}

From the data, we reconstruct the complex acoustic field as:
\begin{equation}
    \Psi_j (\mathbf{\rho}_j +\mathbf{R}\, ,\nu) = \mathcal{A}(\mathbf{\rho}_j + \mathbf{R}\, , \nu) ~ e^{i\phi(\mathbf{\rho}_j + \mathbf{R}\, , \nu)}
\end{equation}
($j = $ A, B, C, H) where $\mathcal{A}$ and $\phi$ denote the local amplitude and phase, respectively, at frequency $\nu$ and lattice site $\mathbf{\rho}_j + \mathbf{R}$ with basis vectors $\mathbf{\rho}_j$ and lattice vectors $\mathbf{R}$. To extract the experimental dispersion relation, we project the real-space field onto a plane-wave basis. By applying a two-dimensional discrete Fourier transform,
%\begin{equation}
%    \Psi_j (\mathbf{k}, \nu) = \sum_{\mathbf{R}} ~ \Psi_j (\mathbf{\rho}_j + \mathbf{R}, \nu) ~ e^{-i \mathbf{k} \cdot \mathbf{R}}
%\end{equation}
\begin{equation}
    \Psi_j (\mathbf{k}, \nu) = \sum_{\mathbf{R}} ~ \Psi_j (\mathbf{\rho}_j + \mathbf{R}, \nu) ~ e^{-i \mathbf{k} \cdot (\mathbf{\rho}_j + \mathbf{R})}
\end{equation}
we obtain the momentum-space representation of the lattice response. The resulting dataset $\Psi_j (\mathbf{k}, \nu)$ and the spectral intensity $I_\mathbf{k} = \sum_j| \Psi_j (\mathbf{k}, \nu) | ^ 2$ (see \cref{fig:figure1}(c)) allow us to measure the experimental band structure and resolve the underlying sublattice information.
%
%
%
\iffalse
From the data, we reconstruct the complex acoustic field as:
\begin{equation}
    \Psi (\mathbf{R}, \nu) = \mathcal{A}(\mathbf{R}, \nu) ~ e^{i\phi(\mathbf{R}, \nu)}
\end{equation}
where $A$ and $\phi$ denote the local amplitude and phase, respectively, at frequency $\nu$ and lattice site $\mathbf{R} = (x, y)$. To extract the experimental dispersion relation, we project the real-space field onto a plane-wave basis. By applying a two-dimensional discrete Fourier transform,
\begin{equation}
    \Psi (\mathbf{k}, \nu) = \sum_{\mathbf{R}} ~ \Psi (\mathbf{R}, \nu) ~ e^{-i \mathbf{k} \cdot \mathbf{R}}
\end{equation}
we obtain the momentum-space representation of the lattice response. The resulting dataset $\Psi_j (\mathbf{k}, \nu)$ and the spectral intensity $I_\mathbf{k} = \sum_j| \Psi_j (\mathbf{k}, \nu) | ^ 2$ ($j = $ A, B, C, H) (see \cref{fig:figure1}(c)) allow us to measure the experimental band structure and resolve the underlying sublattice information. % characterstics % coherent (interference) $| \sum \Psi_i | ^ 2$ and incoherent (interference-free) $\sum | \Psi_i | ^ 2$
% $I (k_x, k_y, \nu, I_\mathbf{k})$
%
%
%
\fi
%
%
%
To interpret the experimental results, we used a minimal tight-binding model that captures the kagome band structure. 
%We assume isotropic nearest-neighbor hopping $t$ among sublattices A, B, and C, which is overlaid on the measurements as solid lines (see \cref{fig:figure1}c). To accurately reproduce the observed isofrequency surfaces, we furthermore account for hopping to the hollow site (H), $t_\textrm{H}$ (see \cref{fig:figure1}d,e).
Details of the model are provided in the Supplemental Material.
\\
\\
The reconstructed acoustic band structure is shown along a high-symmetry $\Gamma \, \textrm{-} \, \mathrm{M} \, \textrm{-} \, \mathrm{K} \, \textrm{-} \, \Gamma$ line and as constant frequency cuts in \ref{fig:figure1}(c) and (d), respectively. Overall, the data is in good qualitative agreement with the TB model (black line in \ref{fig:figure1}(c) and \ref{fig:figure1}(e)), with the key characteristic features of the kagome lattice  being directly evident: two van Hove singularities at $5.5\,\mathrm{kHz} ~ (-2\,t)$ and $8.0\,\mathrm{kHz} ~ (0\,t)$, a Dirac cone near $6.7\,\mathrm{kHz} ~ (-1\,t)$, and a flat band at $\approx 10.9\,\mathrm{kHz} ~ (+2\,t)$. TB values are given with respect to $\epsilon_0$. Fitting these spectral features yields $t = 1.4\,\mathrm{kHz}$ and $\epsilon_0 = 8.05\,\mathrm{kHz}$. At low frequencies, the measured dispersion becomes linear and deviates from the quadratic dispersion in the TB approximation. This behavior reflects the long-wavelength limit $(\lambda \gg a)$ of sound propagation, where the acoustic waves perceive an effective continuum rather than the discrete lattice.
In the TB case, the electron-like band dispersion at small $k$ becomes quadratic $E(\mathbf{k}) \approx \frac{\hbar^2 \mathbf{k}^2}{2m^\ast}$, with an effective quasiparticle mass $m^\ast$. The acoustic spectrum, in turn, is $\omega(\mathbf{k})=2\pi f(\mathbf{k}) \approx c^\ast \mathbf{k}$. Here, $c^\ast$ is an effective sound velocity for which we find $c^\ast \approx 256\,\mathrm{m/s}$ in our acoustic lattice.
Around the zone boundaries (with small wavelengths $(\lambda \approx a)$), the acoustic excitation spectrum clearly reflects the prototypical single-orbital kagome band dispersion. That is, the acoustic spectrum can be described by the (electronic) TB model.
Although we will focus on these distinct kagome features in the following, we should still draw attention to an additional low-intensity feature with almost vanishing dispersion at $\nu \approx 5.5\,\mathrm{kHz}$. This is not captured by the standard three-band model, but we instead attribute it to the hollow site, which can be treated as a fourth, weakly-coupled resonator. Further discussion on this is provided in the SM, including an adjusted four-band TB model and systematic experiments on different acoustic kagome lattices, all of which support this assumption.
Also note that the intensity distribution $I_\mathbf{k} = \sum_j| \Psi_j (\mathbf{k}, \nu=10.9\,\mathrm{kHz}) | ^ 2$ at the flat band (Fig.~\ref{fig:figure1}(d)) is not completely uniform as expected from the three-site TB model (Fig.~\ref{fig:figure1}(e)). This intensity modulation could also be attributed to a slight 'hopping' to the hollow site as well as to a small but finite dispersion of the flat band around the $\Gamma$ point. The latter in turn may be caused by a finite acoustic damping.
\\
\\
A crucial question that arises is whether the observed flat band is really related to kagome physics, i.e., destructive hopping interference. To this end, we focus on the sublattice character at the flat band in ~\cref{fig:flat-band}. In panel (a) (top), we plot the experimental momentum-space texture of the sublattice character, which appears to be characteristically modulated.
Along the $\Gamma - \mathrm{K}$ directions, the texture is dominated by a single sublattice, while along $\Gamma - \mathrm{M}$ directions, it involves contributions from two sublattices at a time. Overall, the $k$-space sublattice composition evolves cyclically as $\textrm{A} \rightarrow \textrm{AB} \rightarrow \textrm{B} \rightarrow \textrm{BC} \rightarrow \textrm{C} \rightarrow \textrm{CA} \rightarrow \textrm{A}$. The experimental findings are consistent with the TB model results shown in \cref{fig:flat-band}(a) (bottom). This close agreement is a strong hint that the observed non- (or low)-dispersive feature is a \textit{de facto} property of the kagome lattice.
This is further supported by additional comparison of experimental data and TB model calculations, such as a projection of the eigenstates on the coherent sum $I_\mathbf{k}^{coh} = |\sum_j \Psi_j (\mathbf{k}, \nu) | ^ 2$, shown in the SM.
\\
\\
In \cref{fig:flat-band}(b), the real space wave function amplitude $|\mathcal{A}(x,y)|^2$ at the flat band frequency is depicted, with the measured phase $\phi(x,y)$ indicated by the red-blue color code (see legend below). The speaker is placed directly in the middle of the shown area, as indicated. As expected, the excited mode is localized, and the acoustic wave propagates only along the two adjacent honeycomb rings, whereas the amplitude is almost completely suppressed outside these rings. In fact, this is caused by destructive interference arising from a pairwise $\pi$ phase shift between adjacent resonators (see alternating red-blue pattern). This is a direct experimental verification of the proposed hopping-interference mechanism in the kagome lattice.

\begin{figure*}
    \centering
    \includegraphics[width=1\linewidth]{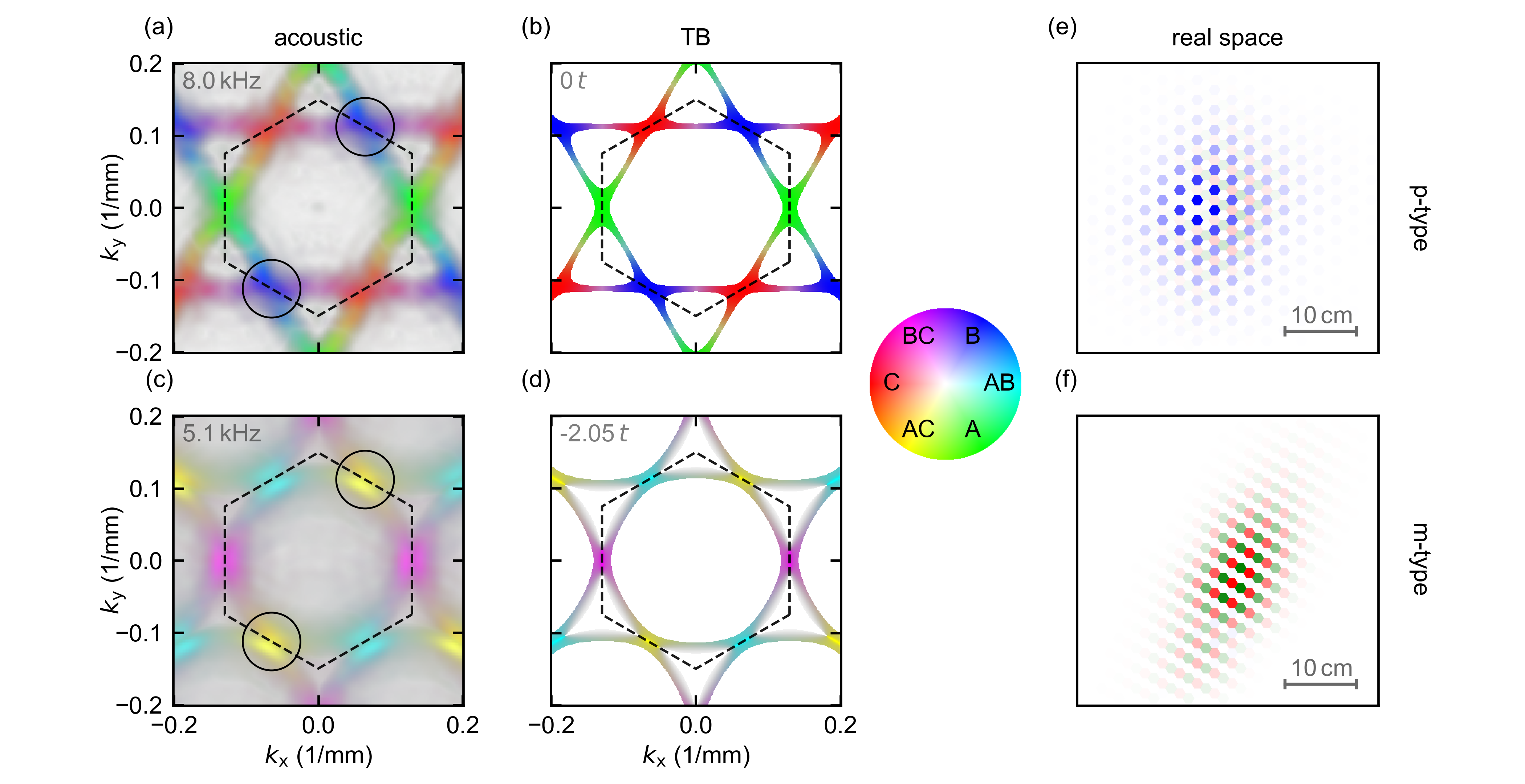}
    \caption{Experimental (left column) isofrequency contours and TB calculated constant energy cuts (middle column). The upper row shows the p-type sublattice character, i.e. only one sublattice contributes at each $\mathrm{M}$-point, of the upper vHS and the bottom row shows the m-type sublattice character, i.e. two sublattices contribute at each $\mathrm{M}$-point, of the lower vHS. The circles in the measured panels show $\sigma$ of the Gaussian that was used to select the area in $k$-space that was Fourier back transformed. The resulting Fourier filtered real-space plots of the amplitude can be seen in the right column. Both plots show that the wave only exists on the sublattice corresponding to the sublattices at the selected $\mathrm{M}$-point.}
    \label{fig:vHS}
\end{figure*}

% Sublattice-resolved isofrequency contours and flat-band localization. (a) Measured (\textit{upper row}) isofrequency contours and corresponding tight-binding calculations (\textit{lower row}) at the m- and p-type vHSs, as well as the flat band (see column labels). (a, b) At the m-type vHS (lower band edge), two sublattices contribute at each M point, while (c, d) the p-type vHS (upper band edge) is characterized by a single-sublattice contribution. The flat-band isofrequency contour features complex direction-dependent m- and p-type sublattice contributions (note the color wheel). (g) Real-space phase and amplitude map at the flat-band frequency. The map demonstrates sublattice interference on the kagome net, characterized by a $\pi$ phase difference between the A, B, and C sublattices and vanishing amplitude at the hollow sites.

Having established the sublattice characteristics of the flat band, we will now focus on the two vHS at $5.5\,\mathrm{kHz} ~ (-2\,t)$ and $8.0\,\mathrm{kHz} ~ (0\,t)$ (compare \cref{fig:figure1}(c,d)). 
In \cref{fig:vHS}(a-d), we show their momentum-space sublattice textures, determined experimentally (panels (a,c)) and using the TB model (panels (b,d)). As expected from previous model considerations \cite{Kiesel2012, Kiesel2013, Wang2013, Schwemmer2024} and in remarkable agreement between the measured $k$-space maps and the TB results, one finds a different sublattice texture in the two vHS. Near the $\mathrm{M}$-points, the lower-frequency vHS exhibits mixed-type sublattice characteristics involving alternating sublattice pairs, whereas the higher-frequency vHS is pure-type and dominated by a single sublattice.
%To illustrate the mixing of the former, the Supplemental Material explores the sublattice-resolved spectral weight, where each sublattice is involved in four of the six $\mathrm{M}$-points. 
The two singularities exhibit a complementary distribution, i.e., the sublattice contributions in the pure-type case appear precisely at those $\mathrm{M}$-points that remain inactive in the mixed-type case (see Supplemental Material). Collectively, all three sublattices contribute, thereby restoring the full symmetry of the kagome lattice across the two singularities. To the best of our knowledge, this is the first experimental demonstration of the characteristic sublattice texture at the vHS of a kagome system.\\
%
%We furthermore selected 2D Gaussian-like weight masks
%\begin{equation}
%    w(k_\mathrm{x},k_\mathrm{y}) = %A\cdot\mathrm{exp}\left(\frac{(k_\mathrm{x}-%k_\mathrm{x,0})^2+(k_\mathrm{y}-k_\mathrm{y,0})^2}%{\sigma^2}\right)
%\end{equation}
%
%with $\sigma$ indicated by the circles in \cref{fig:vHS} around a symmetry-equivalent pair of opposite $\mathrm{M}$-points in the isofrequency cuts of the two vHSs, enabling a Fourier-filtered reconstruction of the original measurements. 
%
%
After imaging the momentum-space sublattice character, we will now focus on the real-space distribution of the wave function. Specifically, we here use a Fourier-filtered reconstruction of the original measurements by putting weight masks at specific $k$-points in the isofrequency cuts of the two vHS (circles in \cref{fig:vHS}(a,c)). Further details about the procedure as well as the complete Fourier transform can be found in the SM. The Fourier-filtered wave function amplitudes are depicted in \cref{fig:vHS}(e,f).
At these specific regions, the spectral weight is clearly dominated by contributions from sublattice B for the p-type (higher) vHS, and from sublattices A and C for the m-type (lower) vHS. The resulting real-space amplitude maps in \cref{fig:vHS}(e) and (f) encode these momentum-space structures, respectively. We observe two distinct features: First, the intensity is strictly confined to the expected sublattices, with sublattice B lighting up for the p-type vHS and sublattices A and C for the m-type vHS. Second, the spatial patterns form localized wave packet envelopes decorated with interference fringes. This is particularly evident in the m-type vHS, where the active A and C sublattices form discrete chains oriented at $150^\circ$ relative to the horizontal axis, each separated by distinct corridors of vanishing intensity. These envelopes directly reflect the $\mathrm{M}$-point band topology within the selected regions, with their spatial profile and off-center shifts governed by the finite momentum width and slight experimental asymmetries in the spectral weight. In this way, the observed wavefront patterns and sublattice-selective intensity distributions provide a real-space manifestation of the underlying vHS eigenmode structure in $k$-space.
%While this outcome may look trivial, it is in fact this is the first real-space demonstration of the sublattice characteristics of the kagome band structure.
While this outcome may appear straightforward, it represents the first direct, real-space demonstration of the sublattice-polarized characteristics inherent to the kagome band structure.

Our results demonstrate the successful realization of an acoustic kagome lattice with full sublattice resolution of the underlying band characteristics. This level of insight is enabled by the combined access to the real-space wave function and its associated momentum-space texture. This requires access not only to the wave function amplitude but also to its phase, which we realize here in our acoustic lattice.
We find that the lower vHS is of mixed type, while the upper is pure. This particularly important aspect is expected from (previous) tight-binding calculations, and we could image this here experimentally in great detail.
Although kagome-related quantum effects are, of course, found exclusively in electronic systems, we show that their underlying sublattice physics results uniquely from wave propagation in the kagome lattice.
In addition to the saddle-point vHS, the flat band likewise exhibits a characteristic $k$-dependent redistribution of spectral weight across the three sublattices. The isofrequency contours reveal this sublattice character to be highly anisotropic, exhibiting pure and mixed sublattice response along the $\mathrm{K} \, \textrm{-} \, \Gamma \, \textrm{-} \, \mathrm{K}$  and $\mathrm{M} \, \textrm{-} \, \Gamma \, \textrm{-} \, \mathrm{M}$ directions, respectively. This has, to the best of our knowledge, not been considered before, and may act as a useful indicator for kagome-originating flat bands in complex, multi-orbital, electronic kagome materials, in which flat bands may also arise from 'trivially'-localized $d$- or $f$-orbitals.\\

\section{ACKNOWLEDGMENTS}

We acknowledge our former group member Felix Hoffmann for his pioneering work on acoustic lattices and Prof. Dr. Ren\'{e} Matzdorf (Universit\"at Kassel) for inspiring this work. This project was funded by the Deutsche Forschungsgemeinschaft (DFG) through the W\"urzburg-Dresden Cluster of Excellence \textit{ctd.qmat} (EXC 2147, Project ID 390858490) and SFB1170 'ToCoTronics' (Project B07).

%Felix Hofmann, René Matzdorf
%Funding: Cluster, SFB1170 A01

% This work was funded by the Deutsche Forschungsgemeinschaft (DFG, German Research Foundation) -- Project-ID 258499086 -- SFB 1170 (projects C04, C06 and A01) -- SFB 1143 (project-ID 247310070) and the W\"{u}rzburg-Dresden Cluster of Excellence on Complexity and Topology in Quantum Matter ct.qmat (EXC 2147, Project No. 390858490). The authors acknowledge ALBA for beamtime provision under experimental proposal ID 2019093897. We also thankfully acknowledge HZB for the allocation of synchrotron radiation beamtime and the financial support.

\bibliographystyle{apsrev4-1}
\bibliography{Kagomev2}

\end{document}